\documentclass{PoS}
\title{Jet measurements in pp, p--Pb and Pb--Pb collisions with ALICE
  at the LHC}
\ShortTitle{Jets in pp, p--Pb, and Pb--Pb using ALICE}
\author{\speaker{S. K. Prasad}\thanks{On behalf of the ALICE Collaboration.}\\
  Centre for Astroparticle Physics and Space Science, Bose
  Institute, Kolkata, 700091 (INDIA)\\
  E-mail: \email{sprasad@cern.ch}}
\abstract{
We present a systematic study of jet measurements in pp, p--Pb and Pb--Pb
collisions using the ALICE detector at the LHC.
Jet production cross sections are measured in pp collisions at
$\sqrt{s}$ = 2.76 and 7~TeV, in p--Pb collisions at $\sqrt{s_{\rm NN}}$ = 5.02~TeV and in Pb--Pb
collisions at $\sqrt{s_{\rm NN}}$ = 2.76~TeV. Jet shape observables and fragmentation
distributions are measured in pp collisions at 7~TeV. Jets are reconstructed at midrapidity in a
wide range of transverse momentum using sequential recombination jet
finding algorithms ($k_{\rm T}$, anti-$k_{\rm T}$, and SISCone) with
several values of jet resolution parameter $R$ in the range 0.2 --
0.6. Measurements are compared to Next-to-Leading Order (NLO)
perturbative Quantum Chromodynamics (pQCD) calculations and
predictions from Monte Carlo (MC) event generators such as PYTHIA,
PHOJET and HERWIG. Jet production cross sections are well reproduced
by NLO pQCD calculations in pp collisions at $\sqrt{s}$~=~2.76~TeV. MC
models could not explain the jet cross sections in pp collisions at
$\sqrt{s}$ = 7 TeV, whereas jet shapes and fragmentation
distributions are rather well reproduced by these models. The jet
nuclear modification factor $R_{\rm pPb}$ in p--Pb collisions is found to be consistent
with unity indicating the absence of large modifications of the
initial parton distribution or strong final state effects on jet
production, whereas a large jet suppression is observed in Pb--Pb
central events with respect to peripheral events indicating
formation of a dense medium in central Pb--Pb events.
}
\FullConference{}

\begin{document}
\section{Introduction}
In high energy hadronic or nuclear collisions, hard (large momentum
transfer $Q^{\rm 2}$) scattered partons (quarks and gluons) fragment
and hadronize, resulting in a collimated shower of particles known as
a jet~\cite{refJetDef}. 
Jet measurements in pp collisions provide a test of perturbative and
non-perturbative aspects of jet production and fragmentation as
implemented in the MC models, and form a baseline for similar measurements in
nucleus--nucleus (A--A) and proton--nucleus (p--A) collisions. 
In A--A collisions an energetic parton
while passing through the produced medium loses energy via induced gluon
radiation and elastic scattering. Jet studies in A--A collisions in
comparison to pp allow a better understanding of the medium induced
modifications in the fragmentation of hard scattered
partons and energy loss mechanisms~\cite{refJetQuenching1, refJetQuenching2}, whereas similar studies in
p--A collisions potentially reveal the effects of (cold) nuclear matter
(CNM). In this paper we present
results of jet measurements obtained using ALICE detector in pp
collisions at $\sqrt{s}$ = 2.76, 7~TeV, in p--Pb collisions at
$\sqrt{s_{\rm NN}}$ = 5.02~TeV and in Pb--Pb collisions at
$\sqrt{s_{\rm NN}}$ = 2.76~TeV.
\section{Data sample, event selection, track selection, and jet reconstruction}\label{secJetRec}
The data used in this analysis were collected during the LHC run in
2010 for pp collisions at 7~TeV, in 2011 for pp collisions at 2.76~TeV, in fall of 2010 for Pb--Pb
collisions, and in the beginning of 2013 for p--Pb
collisions using the ALICE detector~\cite{refAliceExpt1, refAliceExpt2}.
Minimum bias events are selected based on information from
Silicon Pixel Detector (SPD)~\cite{refSPD} and V0~\cite{refV0}
detectors (V0A, V0C)~\cite{refFullJetPP, refChJetPP, refChJetPPb,
  refChJetPbPb}. The Electromagnetic
Calorimeter (EMCal)~\cite{refEmcal} is used in addition to select EMCal triggered
events~\cite{refFullJetPP}. 
Information from the Time Projection Chamber (TPC)~\cite{refTPC} and Inner Tracking System
(ITS)~\cite{refSPD} are used to select charged tracks using a hybrid
approach as discussed in~\cite{refFullJetPP}. The
reconstruction of neutral particles is performed using the
Electromagnetic Calorimeter (EMCal)~\cite{refFullJetPP}.
Charged tracks with transverse momentum 
$p_{\rm T}^{\rm track}~\textgreater$ 0.15 GeV/$c$ at midrapidity ($\mid\eta^{\rm
  track}\mid~\textless$ 0.9) are used as input to jet
reconstruction. In addition, for jets including neutral particles,
EMCal clusters with energy greater than 0.3 GeV/$c$ are considered.
Jets are reconstructed using the infrared collinear safe sequential
recombination algorithm anti-$k_{\rm T}$~\cite{refAntiKt}. In p--Pb and
Pb--Pb collisions, the $k_{\rm T}$~\cite{refKt1, refKt2} algorithm is
used for the estimation of background. Jets are reconstructed with
several values of the resolution parameter $R$ in the range 0.2 -- 0.6. Jets reconstructed using charged particles only as input
are referred to as {\it `charged jets'} whereas jets reconstructed using both
charged and neutral particles as input are known as {\it `full jets'}
hereafter.
\section{Correction for detector effects and background}\label{secCorrection}
Measured distributions are corrected for the instrumental effects and presented
at particle level. Corrections for the instrumental effects, such as limited track
reconstruction efficiency and finite momentum resolution, are
performed using the unfolding
techniques~\cite{refBayesianUnf, refSVDUnf} for jet cross sections
whereas jet shape and fragmentation observables are corrected using
a bin-by-bin technique. A full detector simulation is performed using
the PYTHIA~6.425~\cite{refPythia6} event generator and GEANT3~\cite{refGeant3}
particle transport package. All
observables are also corrected for the
contamination from the secondary particles~\footnote{Particles
  produced by weak decays and interactions of primary particles with
  detector material and beam pipe.} and background~\footnote{Particles
  in an event which are not produced directly by hard scattering of
  partons.}. The method for
estimation and correction of background is however different in pp to that in
p--Pb and Pb--Pb collisions (see~\cite{refFullJetPP, refChJetPP,
  refChJetPPb,refChJetPbPb} for details). In case of p--Pb and Pb--Pb
events, region to region fluctuations of the estimated average
background density arising due to fluctuations in the particle multiplicity and momentum,
elliptic flow etc., are considerable. Background fluctuations are corrected for on an
statistical basis using unfolding techniques (see Sec.~2
of~\cite{refChJetPbPb}). 
The corrected results are compared to that obtained from various MC
event generators e.g. PYTHIA (tune Perugia-0, Perugia-2011, AMBT1), HERWIG,
PHOJET and NLO pQCD calculations.
\section{Results}
\subsection{Jet measurements in pp collisions}
\begin{figure*}[ht]
  \begin{center}
    \includegraphics[scale=0.28]{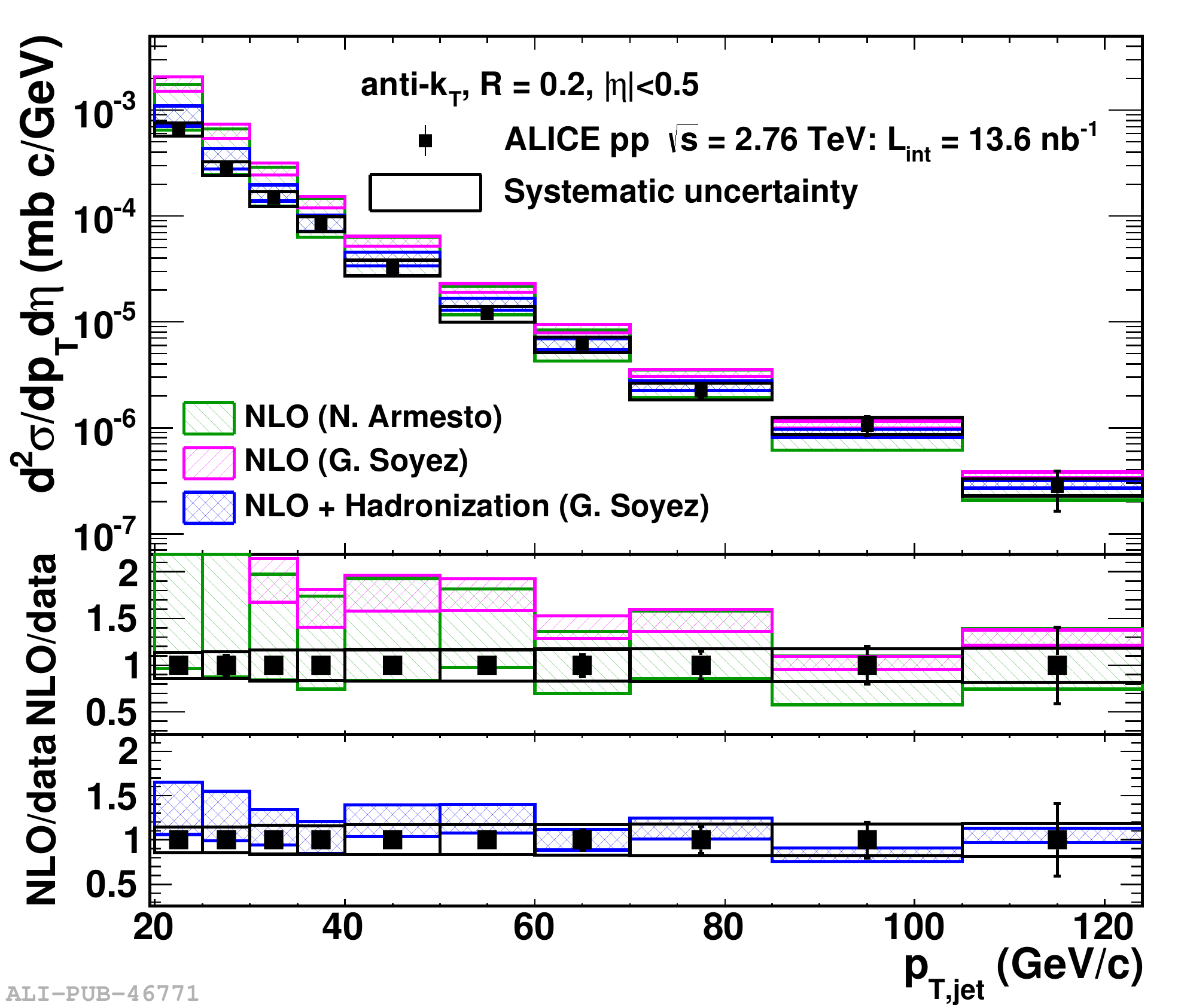}
    \includegraphics[scale=0.28]{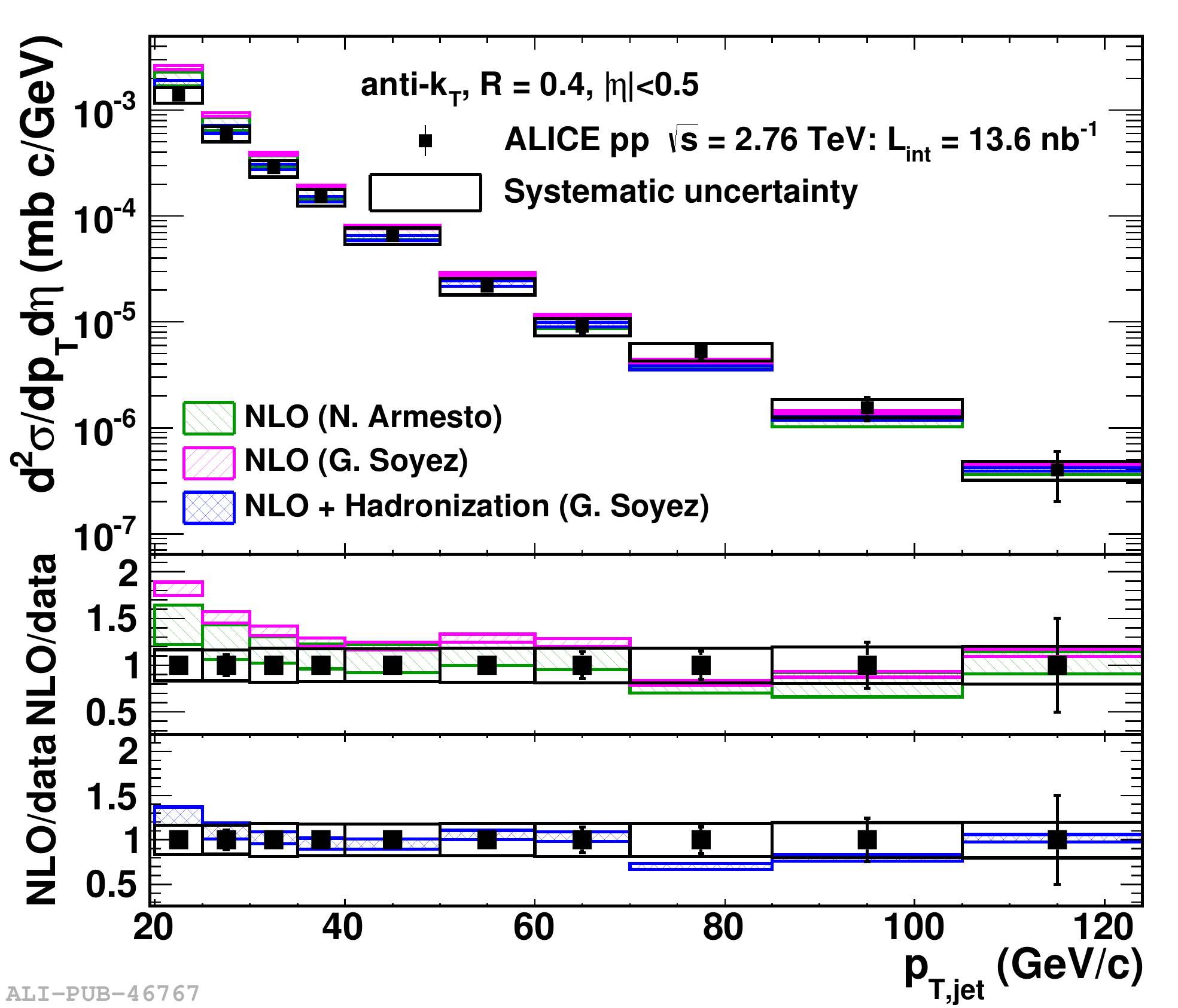}
    \caption{(Color online) Full jet production cross sections  
      as a function of jet $p_{\rm T}$ compared to NLO pQCD  
      calculations in pp collisions at $\sqrt{s}$~=~2.76~TeV for jets  
      reconstructed with $R$~=~0.2~(left) and 0.4~(right).}
    \label{FigXsecFull}
  \end{center}
\end{figure*} 
\begin{figure*}[ht]
  \begin{center}
    \includegraphics[scale=0.26]{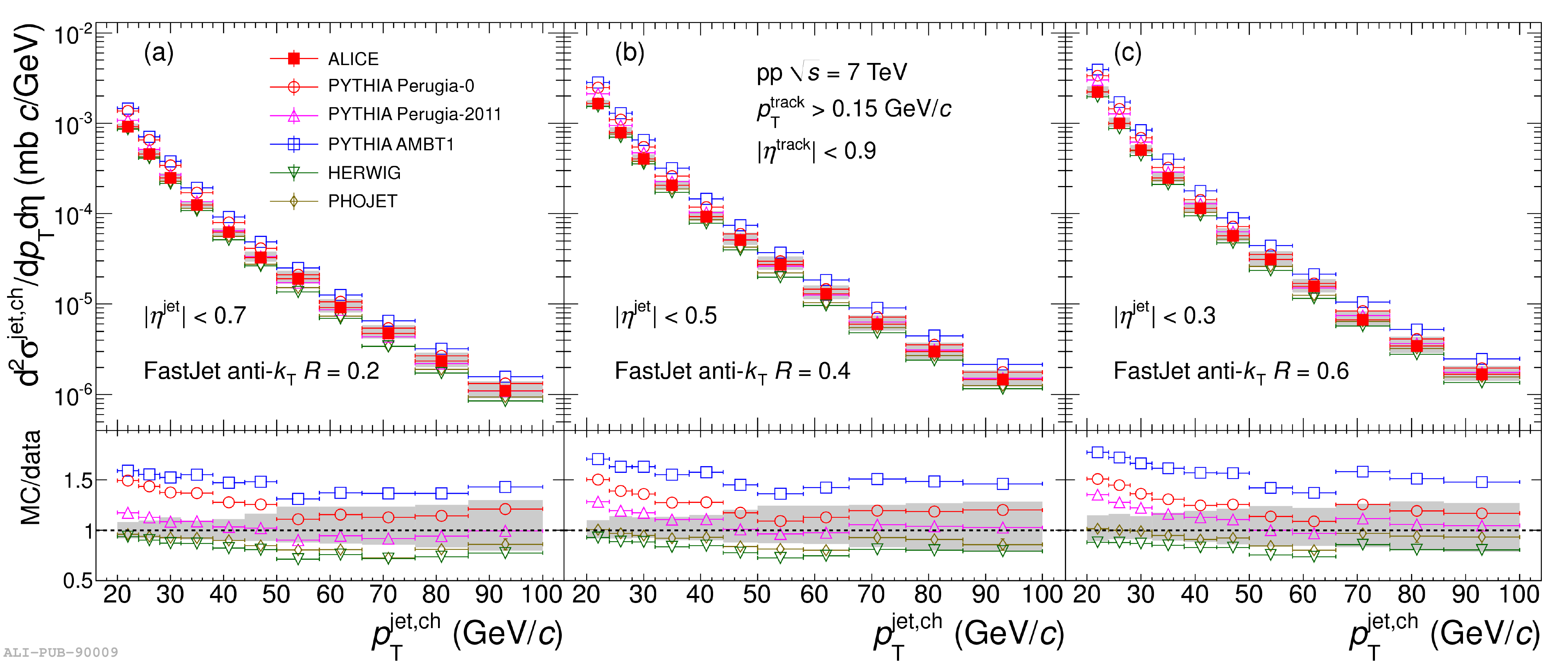}
    \caption{(Color online) Charged jet cross sections as a function
      of jet $p_{\rm T}$ compared to MC models in pp collisions at
      $\sqrt{s}$~=~7~TeV for jets reconstructed with $R$~=~0.2~(left), 0.4~(middle) and 0.6~(right).}  
    \label{FigXsecCharged}
  \end{center}
\end{figure*}
The full jet production cross
sections as a function of jet $p_{\rm T}$ compared to NLO pQCD
calculations~\cite{refNLOpQCD} are shown in Fig.~\ref{FigXsecFull}~\cite{refFullJetPP}
for pp collisions at $\sqrt{s}$ = 2.76~TeV for jets reconstructed with $R$ = 0.2 (left) and 0.4 (right).
The NLO calculations reproduce the full jet cross sections reasonably
well when hadronization effects are included. 
The charged jet cross sections compared to MC predictions, are shown in Fig.~\ref{FigXsecCharged}
for pp collisions at $\sqrt{s}$~=~7~TeV for $R$~=~0.2~(left),
0.4~(middle) and 0.6~(right). None of the models can explain the
data in the entire $p_{\rm T}$ range, the discrepancy being larger for
larger $R$. The jet
shape observables as defined by the  radial transverse momentum density distributions about the jet
axis as a function of distance `$r$', jet constituents multiplicity and
average radius containing 80\% of jet $p_{\rm T}$ as a function of
leading jet $p_{\rm T}$, and the
fragmentation distributions are, however, in general reasonably well reproduced by
these models (figures not shown, see~\cite{refChJetPP}).
\subsection{Results from p--Pb and Pb--Pb collisions}
\begin{figure*}[ht]
  \begin{center}
    \includegraphics[scale=0.6]{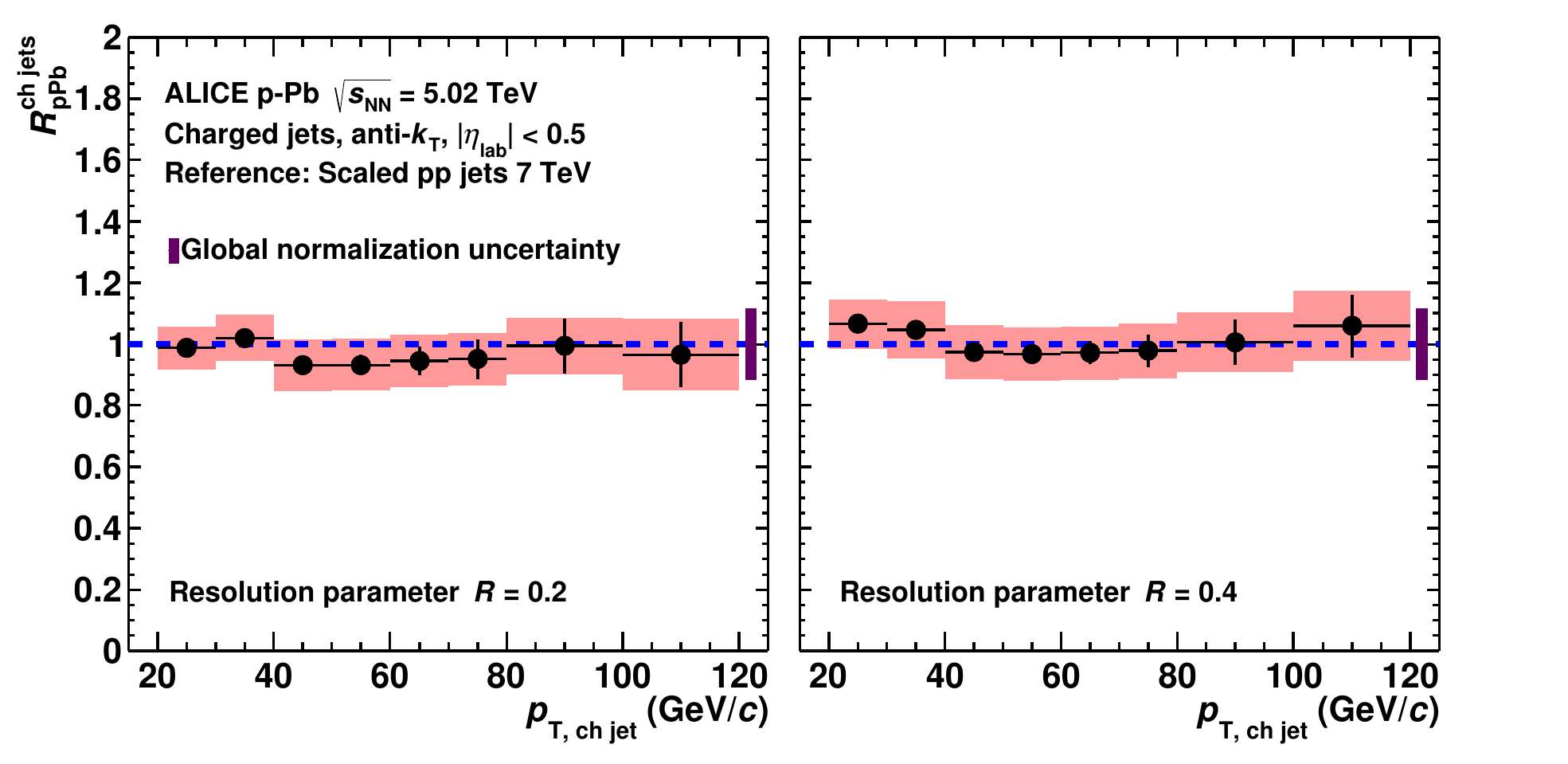}
    \caption{(Color online) Jet nuclear modification factors ($R_{\rm pPb}$) in p--Pb collisions 
      at $\sqrt{s_{\rm NN}}$~=~5.02~TeV for charged jets reconstructed with
      $R$~=~0.2~(left) and 0.4~(right).}
    \label{FigRpPb}
  \end{center}
\end{figure*}
The jet nuclear
modification factor $R_{\rm pPb}$~\footnote{$R_{\rm pPb}$ is defined
  as the ratio of
  $p_{\rm T}$ spectra in p--Pb normalized by the nuclear overlap function $\langle T_{\rm AA}\rangle$
  obtained from Glauber model and pp cross section extrapolated to
  $\sqrt{s_{\rm NN}}$ = 5.02~TeV.}
is shown in Fig.~\ref{FigRpPb} for p--Pb collisions at $\sqrt{s_{\rm
    NN}}$~= 5.02~TeV for charged jets reconstructed with $R$ = 0.2 (left) and
0.4 (right). It is found to be
consistent with unity in the measured $p_{\rm T}$ range indicating
the absence of large modifications of the initial parton
distribution or strong final state effects on jet production.
\begin{figure*}[ht]
  \begin{center}
    \includegraphics[scale=0.30]{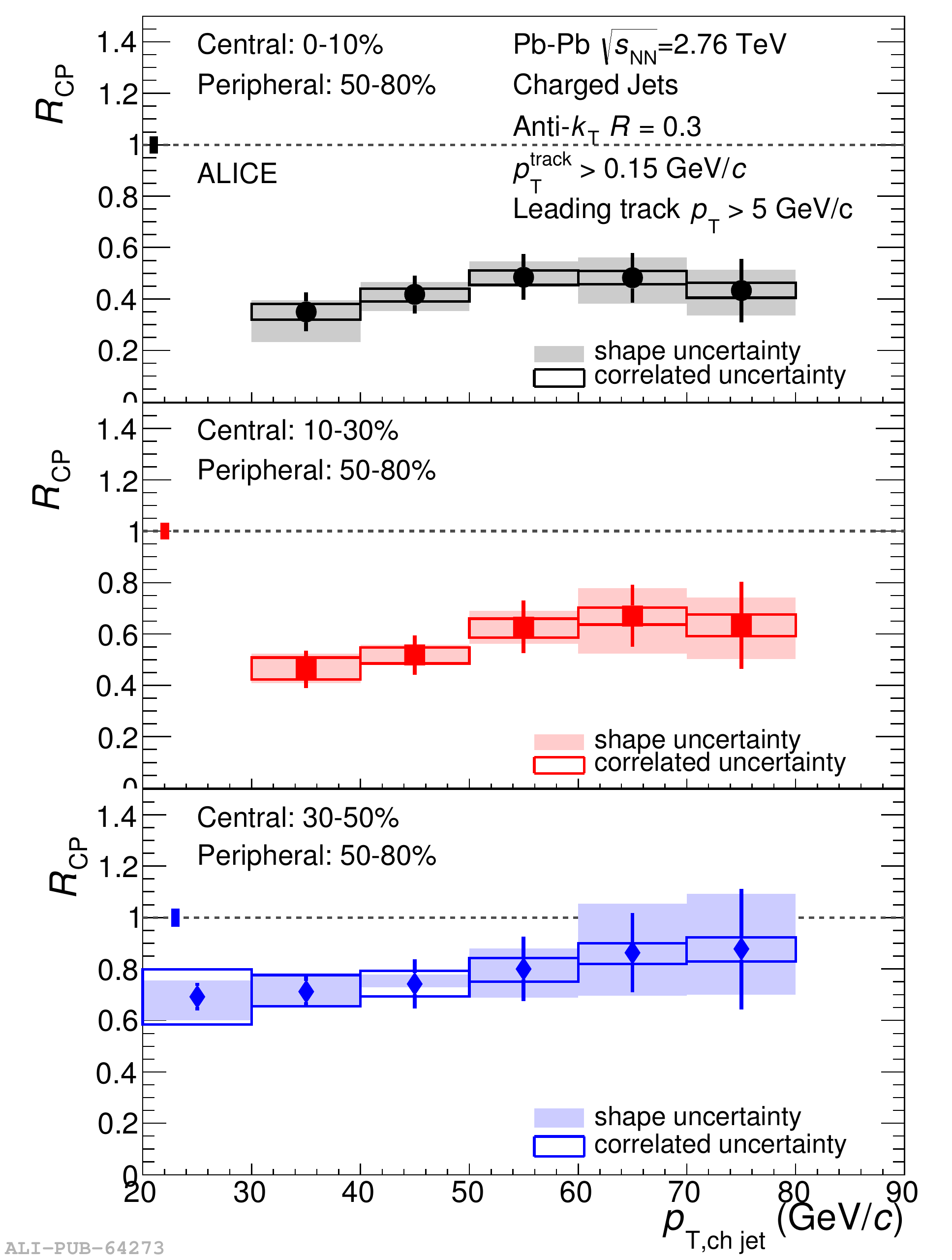}
    \caption{(Color online) Charged jet nuclear modification factors ($R_{\rm 
        CP}$) in Pb--Pb collisions 
      at $\sqrt{s_{\rm NN}}$~=~2.76~TeV for 0--10\% (top), 10--30\% (middle)
      and 30--50\% (bottom) centrality classes.}
    \label{FigRaa}
  \end{center}
\end{figure*}
The charged jet nuclear modification factor, $R_{\rm CP}$~\footnote{$R_{\rm CP}$ is defined as
the ratio
of jet $p_{\rm T}$ spectra in central and peripheral Pb--Pb collisions
normalized by $\langle T_{\rm AA}\rangle$
for each centrality class.} is
shown in Fig.~\ref{FigRaa} as a function of jet $p_{\rm T}$ for
three centrality bins for Pb--Pb collisions at
$\sqrt{s_{\rm NN}}$~=~2.76~TeV. A large
jet suppression is observed in most central
(0--10\%) Pb--Pb collisions indicating the formation of a dense medium in
such collisions. It is found to be centrality and $p_{\rm T}$ dependent.
\begin{figure*}[ht]
  \begin{center}
    \includegraphics[scale=0.32]{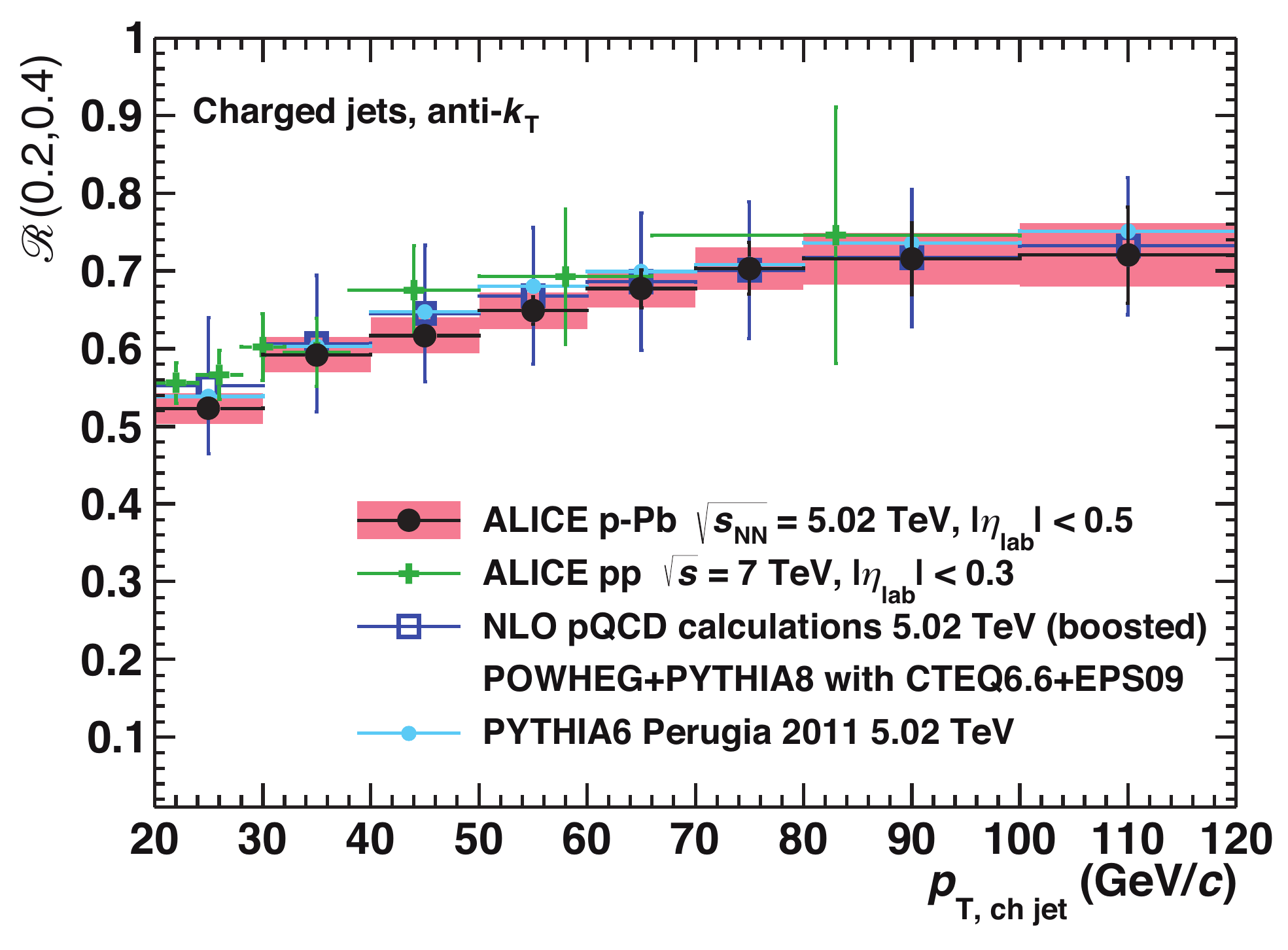}
    \includegraphics[scale=0.35]{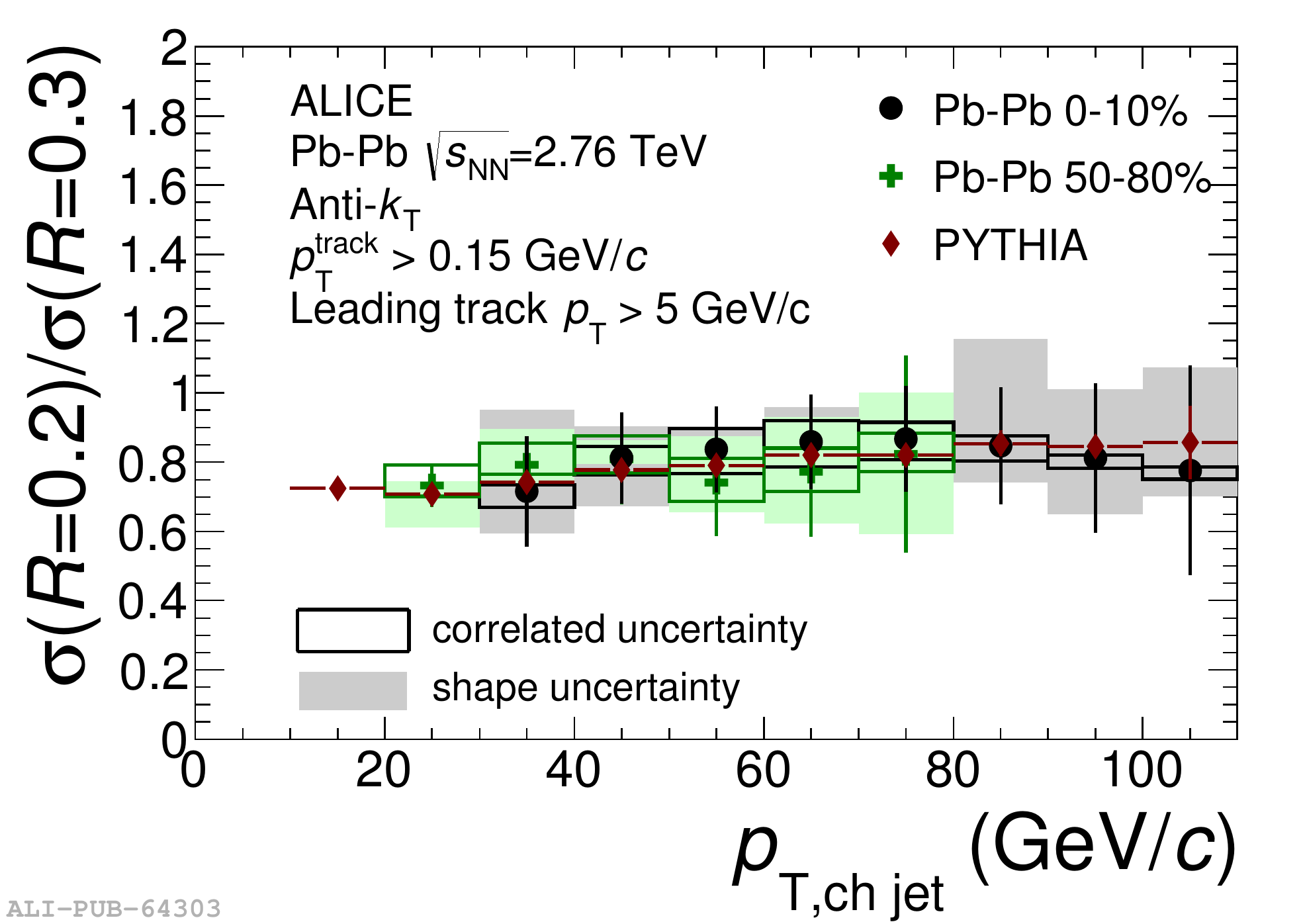}
    \caption{(Color online)The ratios of jet spectra measured with $R$ =
      0.2 to that obtained with larger $R$ (0.4 for p--Pb and 0.3 for
      Pb--Pb) as a function of jet $p_{\rm T}$ in p--Pb~(left) and Pb--Pb~(right)
      collisions at 5.02 and 2.76 TeV respectively.}
    \label{FigXsecRatios}
  \end{center}
\end{figure*} 
The left (right) panel of Fig.~\ref{FigXsecRatios} shows ratios of spectra (cross sections)
for jets measured with $R$ = 0.2 and 0.4 (0.2 and 0.3) in p--Pb
(Pb--Pb) collisions at $\sqrt{s_{\rm NN}}$ = 5.02~TeV (2.76~TeV) compared to
that obtained in pp collisions (PYTHIA). The ratio of jet spectra is
sensitive to the collimation of particles around the jet axis and
serves as an indirect measure of jet structure. In minimum bias p--Pb collisions the ratio of jet
spectra is found to be compatible with
that in pp collisions, PYTHIA and NLO pQCD calculations, and the cross section ratios in Pb--Pb is
found to be similar for most central and peripheral collisions and
compatible with PYTHIA indicating that the
core of the jet within the measured $R$, remains unmodified in minimum
bias p--Pb, peripheral Pb--Pb and even in most central Pb--Pb collisions.

\section{Summary and conclusions}
We reported jet measurements for pp, p--Pb and Pb--Pb
collisions at various centre-of-mass energies using the ALICE detector. Jets are measured at midrapidity
using the anti-$k_{\rm T}$ jet finding algorithm with several values
of the jet resolution parameter ($R$ in the range 0.2 to 0.6). Full jet cross
sections are well reproduced by NLO pQCD calculations in pp collisions
at $\sqrt{s}$ = 2.76~TeV. None of the MC models under study can explain
the charged jet cross sections in pp collisions at
$\sqrt{s}$ = 7~TeV, however jet shape observables and fragmentation
distributions are rather well reproduced by these models. The jet
nuclear modification factor for minimum bias p--Pb collisions is found
to be consistent with unity whereas a large jet suppression is
observed for central Pb--Pb events with respect to peripheral events
indicating the presence of a dense medium in these collisions. The jet
spectra (or cross section) ratios indicate that the core of the jet
remains unmodified even in the most central Pb--Pb collisions.

\end{document}